\documentclass[secnumarabic,amssymb, nobibnotes, aps, prd, 11pt, abstract]{revtex4-2}
\usepackage{tocloft}
\usepackage[dvips]{graphicx}
\usepackage{amsmath}
\usepackage{subcaption}
\usepackage{hyperref}
\usepackage{filecontents}
\hypersetup{colorlinks,linkcolor={red},citecolor={blue},urlcolor={red}}
\usepackage{setspace}
\usepackage{xcolor}
\pagecolor{white}
\usepackage{float}

\usepackage{multirow}
\color{black}

\begin{document}

\singlespacing

	\title{Lyapunov Exponents, Phase Transitions, and Chaos Bound of ModMax AdS Black Holes}%

	\author{Gorima Bezboruah$^1$}
	
	\email{$gorimabezboruah@gmail.com$}
	
	\author{Mozib Bin Awal$^1$}
	
	\email{$mozibawal@gmail.com$}

	\author{Prabwal Phukon$^{1,2}$}
	\email{$prabwal@dibru.ac.in$}
	
	\affiliation{$^1$Department of Physics, Dibrugarh University, Dibrugarh, Assam,786004.\\$^2$Theoretical Physics Division, Centre for Atmospheric Studies, Dibrugarh University, Dibrugarh,Assam,786004.\\}

	\begin{abstract}
	
We study the thermodynamic phase transition of ModMax anti-de Sitter (AdS) black holes using Lyapunov exponents of massless and massive particles in unstable circular orbits. Our results demonstrate that the thermal profile of the Lyapunov exponent serves as an efficient probe of the black hole’s phase structure. We calculate the discontinuity in the Lyapunov exponent across the transition and show that it acts as an order parameter, exhibiting a critical exponent $\delta=1/2$ in the vicinity of the critical point.  Furthermore, we explore the violation of the chaos bound, finding that the bound is violated when the horizon radius falls below a threshold value. We also examine how the ModMax parameter and the particle’s angular momentum modify this threshold, revealing their role in controlling the onset of chaos bound violation.
 
\end{abstract}
	
	\maketitle
	
\section{Introduction}\label{sec1}

Black hole thermodynamics is one of the most fascinating areas of research. Since pioneering works of Bekenstein, Hawking, Bardeen  and others \cite{Phys,bekens,Hawking,Hawking2,Bardeen} during the fromative years of black hole thermodynamics, it has remained an active are of research in modern physics. In the subsequent years, further research firmly established that black holes indeed behave as thermodynamic objects leading to the natural question of whether black holes also undergo processes analogous to those in classical thermodynamics. ``Phase Transitions" in black holes is one of such phenomenon that is of significant interest. It was first explored by P.C.W Davies and P Hut in \cite{Davies} and \cite{Hut} respectively. Thermodynamic studies on Anti de Sitter black holes gained particular attention due to Maldacena's proposal of the AdS/CFT correspondence \cite{Maldacena} in $1997$. Another major development in this field was the interpretation of the cosmological constant $\Lambda$ as thermodynamic pressure term in the first law of black hole thermodynamics. Following this, several studies have adopted this ``extended" thermodynamic framework and found very rich and intricate thermodynamic phase structure. Of particular interest among them is the striking resemblance of the phase transition of black holes to that observed in fluid systems \cite{Kubiz,Hawkpage,Cai,Kastor,Dolan,Dolan2,Dolan3,Kubizna,Xu,Xu2,Zhang}. The conventional approach to studying black hole phase transitions has been by analyzing the free energies and specific heats. However, various alternative approaches to study phase transitions in black holes has been gaining popularity lately. The well-established and extensively studied approaches include the geometric analysis of the thermodynamic state space. The most widely used technique in this context is the study of the Ruppeiner geometry \cite{Ruppeiner:2012uc,Miao:2017cyt,Guo:2019oad,Wei:2019yvs,Wang:2019cax,Yerra:2020oph,Yerra:2021hnh}. Another notable approach involves studying the topology of the thermodynamic space \cite{Wu:2022whe,Liu:2022aqt,Fan:2022bsq,Gogoi:2023xzy,Ali:2023zww,Saleem:2023oue,Shahzad:2023cis,Chen:2023elp,Bai:2022klw,Yerra:2022alz,Hazarika:2023iwp} and it has proved to be a powerful tool for understanding phase transitions and critical behaviour in black hole systems. Apart from these purely theoretical approaches, there are studies that have explored the possible connections between phase transitions in black holes and observable astrophysical phenomenon. A few examples of such approaches include analyses of Quasinormal Modes (QNMs) \cite{Liu:2014gvf,Zou:2017juz,Zhang:2020khz,Mahapatra:2016dae,Chabab:2016cem}, studying the behaviour of test particle in circular orbits \cite{Wei:2017mwc,Wei:2018aqm,Zhang:2019tzi} and the characteristics of black hole shadows \cite{Zhang:2019glo,Belhaj:2020nqy}.

Chaos theory is a field developed by physicists and mathematicians to investigate the dynamical behaviour of systems that are highly sensitive to initial conditions. A quantity that is central to the chaos theory is the Lyapunov exponent \cite{lyp} which measures the rate of divergence or convergence of nearby trajectories in phase space over time \cite{lyp2}. Having a positive Lyapunov exponent implies exponential divergence, signifying a chaotic behaviour. On the other hand, a negative Lyapunov exponent implies convergence and stability within the system. Chaotic dynamics has been used in several studies to investigate phase transitions in quantum as well as classical regimes. Some notable examples include the well-known Sachdev-Ye-Kitaev (SYK) model \cite{syk,syk2}, the Dicke model \cite{Dicke}, finite Fermi systems and quantum dots \cite{finite}, as well as long-range coupled oscillator models \cite{coscll}, among others. More recently, Lyapunov exponents have been employed to analyze the chaotic dynamics in the context of general relativity and black hole physics. Several investigations has focused on the motion of particles in static, axisymmetric backgrounds \cite{static}, in the vicinity of rotating and charged Kerr–Newman black holes \cite{kerr,kerr2}, in multi–black hole systems \cite{multi}, as well as in scenarios where quantum gravity effects are incorporated into black hole solutions \cite{qg,qg2}. Maldacena, Shenker and Stanford introduced a universal upper bound, which is now known as the MSS bound, that sets a limit on the rate at which chaos can grow in quantum systems with large degrees of freedom and having a well-defined semiclassical gravity dual \cite{mss}. This bound was expressed in terms of the Lyapunov exponent $\lambda$. In natural units, the MSS bound takes the form $\lambda_L\leq2\pi\tilde{T}$. Following this, other studies have confirmed the validity of this bound in diverse settings, including the dynamics of massive particles near black hole horizons \cite{hori}, and its deep connection to the presence of event horizons \cite{hori2}. Nonetheless, MSS bound violation has also been reported in several contexts \cite{vio,vio2,vio3,vio4}.

Probing the thermodynamic phase structure of black holes using Lyapunov exponents was first attempted in reference \cite{first}. In this work, the authors proposed a conjecture suggesting a connection between Lyapunov exponents and black hole phase transitions. It was demonstrated that the Lyapunov exponent corresponding to massless and massive particles circularly orbiting around black holes exhibit multivalued behaviour when plotted against temperature. The distinct branches of the Lyapunov exponent can be interpreted as representing different thermodynamic phases of black hole. The multivaluedness was found to disappear at a specific critical value of model dependent parameters. It was also shown that the discontinuous change in the Lyapunov exponent across the transition can be identified as an order parameter, with a critical exponent of $1/2$ near the phase transition point. Subsequently, several other works have confirmed and extended the original results \cite{le,le2,le3,le4,le5,le6,le7,le8,Awal,le9,le10}

In this paper, we extend the investigation of the interplay between Lyapunov exponents and black hole thermodynamics to a particular nonlinear electrodynamics (NLED) black hole proposed in ref \cite{mm}, which is commonly known as ModMax NLED black hole. ModMax electrodynamics is an extension of Maxwell theory incorporating conformal invariance and nonlinear field strength corrections and providing a rich platform for exploring the interplay between gravitational dynamics and gauge field nonlinearity. The organisation of the paper is done as follows. In section \ref{sec2} we offer an overview of the methodology used to compute Lyapunov exponents. In section \ref{sec3} we outline the key features and review the thermodynamics of ModMax AdS black holes. In section \ref{sec4}, we calculate the Lyapunov exponent associated with both massless and massive particles in circular orbits around the black hole and probe its thermodynamic phase transition using it. In section \ref{sec5}, we examine the discontinuity in the Lyapunov exponent near the critical point and extract the associated critical exponent. Section \ref{sec6} offers a detailed analysis of the chaos bound and its violation. We summarize our results in section \ref{sec7}

\section{Review of Lyapunov exponents}\label{sec2}
We present a concise summary of the procedure for computing Lyapunov exponents in this section. We consider the dynamics of both massless and massive particles executing unstable circular orbits. We start off by writing the Lagrangian governing the geodesic motion of such particles within the given spacetime background $\theta=\pi/2$
\begin{equation}\label{eq1}
2\mathcal{L}=-f(r)\dot{t}^2+\frac{\dot{r}^2}{f(r)}+r^2\dot{\phi^2}
\end{equation}
where the dot denotes the differentiation with respect to the proper time $\tau$. The canonical momenta corresponding to the generalized coordinates are derived from the Lagrangian using the standard relation $p_q=\frac{\partial\mathcal{L}}{\partial\dot{q}}$, giving us,
\begin{equation}\label{eq2}
p_t=\frac{\partial\mathcal{L}}{\partial\dot{t}}=-f(r)\dot{t}=-E,\; p_r=\frac{\partial\mathcal{L}}{\partial\dot{r}}=\frac{\dot{r}}{f(r)},\; p_\phi=\frac{\partial\mathcal{L}}{\partial\dot{\phi}}=r^2\dot{\phi}=L
\end{equation}
where $E$ and $L$ are the conserved energy and angular momentum of the particle. From equation (\ref{eq2}) we get \begin{equation}\label{eq3}
\dot{t}=\frac{E}{f(r)} \quad \dot{\phi}=\frac{L}{r^2}
\end{equation}
Using the standard definition of Hamiltonian and the relations in (\ref{eq3}), we can write
\begin{equation}\label{eq4}
2\mathcal{H}=-E\dot{t}+\frac{\dot{r}^2}{f(r)}+\frac{L^2}{r^2}=\delta_1
\end{equation}
where $\delta_1 = -1$ corresponds to timelike geodesics and $\delta_1 = 0$ to null geodesics. The effective potential governing the radial motion of the particle is given by,
\begin{equation}\label{eq5}
V_{\text{eff}}=f(r)\left[\frac{L^2}{r^2}+\frac{E^2}{f(r)}-\delta_1\right]
\end{equation}
The angular momentum can be expressed in terms of the effective potential by setting $E=0$ in equation (\ref{eq5}) and substituting the result into equation (\ref{eq4}) allows us to rewrite the Hamiltonian as
\begin{equation}\label{eq6}
\mathcal{H}=\frac{V_{\text{eff}}-E^2}{2f(r)}+\frac{f(r)p^2_r}{2}+\frac{\delta_1}{2}
\end{equation}
Hence the equation of motion in proper time configuration can be expressed as follows
\begin{equation}\label{eq7}
\dot{r}=\frac{\partial\mathcal{H}}{\partial p_r}=f(r)p_r,\quad \dot{p_r}=\frac{\partial\mathcal{H}}{\partial r}=-\frac{V'_{\text{eff}(r)}}{2f(r)}-\frac{f'(r)p^2_r}{2}+\frac{V_{\text{eff}}-E^2}{2f^2(r)}f'(r)
\end{equation}
Here, primes denote derivatives with respect to $r$. Linearizing the equations of motion about the circular orbit at $r_0$, we obtain the corresponding stability matrix $K$  formulated with respect to the coordinate time $t$ as,
\begin{equation}\label{eq8}
K = \begin{pmatrix}
0 & \frac{f(r_0)}{\dot{t}} \\
-\frac{V''_{\text{eff}}(r_0)}{2f(r_0)\dot{t}} & 0
\end{pmatrix}
\end{equation}
The Lyapunov exponent is obtained as the eigenvalue of the stability matrix above
\begin{equation}\label{eq9}
\lambda=\sqrt{-\frac{V''_{\text{eff}}(r_0)}{2\dot{t}^2}}
\end{equation}
here we have dropped the $\pm$ sign before the square root for simplicity.
The instability condition for circular geodesics is given by \begin{equation}\label{eq10}
V'_{\text{eff}}(r_0)=0,\quad V''_{\text{eff}}(r_0)<0
\end{equation}
which also gives us the radius of the unstable circular orbit $r_0$.
Using these two relations in equation (\ref{eq5}) we obtain \begin{equation}\label{eq11}
\frac{E}{L}=\frac{\sqrt{f(r_0)}}{r_0}
\end{equation}
and subatituting this in equation (\ref{eq3}) while using $\delta_1=0$ (for massless particles), we get \begin{equation}\label{eq12}
\dot{t}=\frac{L}{r_0\sqrt{f(r_0)}}
\end{equation}
This leads us to obtain the final expression for Lyapunov exponent (\ref{eq9}) as 
\begin{equation}\label{eq13}
\lambda=\sqrt{-\frac{r^2_0f(r_0)}{2L^2}V''_{\text{eff}}(r_0)}
\end{equation}
In a similar manner, for massive particles ($\delta_1 = 1$), the corresponding relations are
\begin{equation}\label{eq14}
E^2=\frac{2f^2(r_0)}{2f(r_0)-r_0f'(r_0)}\quad \text{and} \quad L^2=\frac{r^3_{0}f'(r_0)}{2f(r_0)-r_0f'(r_0)}
\end{equation}
Hence from equation (\ref{eq3}) we obtain
\begin{equation}\label{eq15}
\dot{t}=\frac{1}{\sqrt{f(r_0)-\frac{1}{2}r_0f'(r_0)}}
\end{equation}
Finally, by applying equation (\ref{eq9}), we arrive at the expression for Lyapunov exponent for massive particles
\begin{equation}\label{eq16}
\lambda=\frac{1}{2}\sqrt{\left[r_0f'(r_0)-2f(r_0)\right]V''_{\text{eff}}}
\end{equation}

\section{ModMax AdS Black Hole Background}\label{sec3}
\begin{figure}[h!]
    \centering
    \begin{subfigure}[b]{0.45\textwidth}
        \centering
        \includegraphics[width=\textwidth]{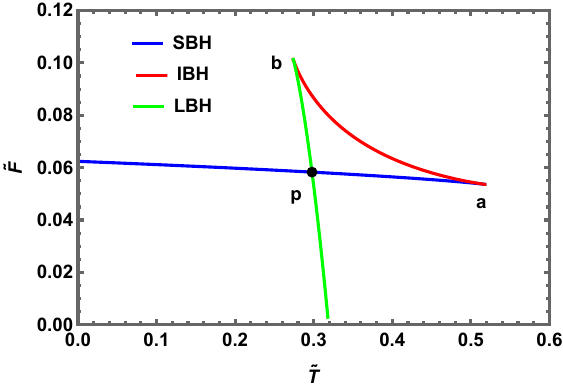}
        \caption{$\tilde{q}=0.08<\tilde{q}_c$ and $\eta=0.5$}
        \label{f2a}
    \end{subfigure}
    
    \vskip\baselineskip  

    \begin{subfigure}[b]{0.45\textwidth}
        \centering
        \includegraphics[width=\textwidth]{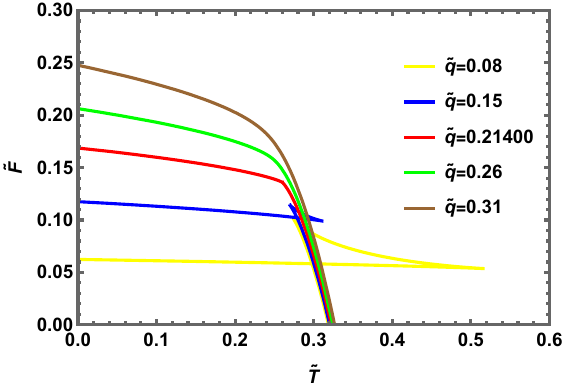}
        \caption{$\eta=0.5$}
        \label{f2b}
    \end{subfigure}
    \hspace{0.8cm}
    \begin{subfigure}[b]{0.45\textwidth}
        \centering
        \includegraphics[width=\textwidth]{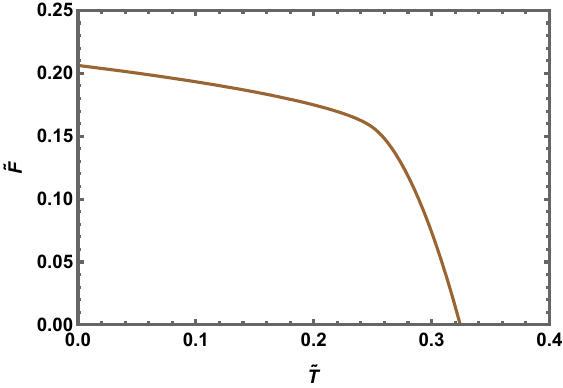}
        \caption{$\tilde{q}>\tilde{q}_c$ and $\eta=0.5$}
        \label{f2c}
    \end{subfigure}

    \caption{Gibbs free energy as a function of temperature}
    \label{f2}
\end{figure}

The action of an anti de Sitter black hole coupled to the ModMax nonlinear electrodynamics field is given by \cite{Sekhmani} \begin{equation}\label{eq17}
\mathcal{I}=\frac{1}{16\pi}\int d^4x\sqrt{-g}\left(\mathcal{R}+6L^{-2}-4\mathcal{L}\right)
\end{equation}
where $\mathcal{R}$ is the Ricci scalar, $L$ is the AdS length $g$ is the determinant of the metric tensor $g_{\mu\nu}$ and $\mathcal{L}$ is the ModMax Lagrangian. This Lagrangian can be expressed as \cite{Kosyakov} \begin{equation}\label{eq18}
\mathcal{L}=\frac{1}{2}\left(\mathcal{S}\cosh{\eta}-\sqrt{\mathcal{S}^2+\mathcal{P}^2}\sinh{\eta}\right)
\end{equation}
where the parameter $\eta$ is defined as a dimensionless intrinsic quantity within ModMax theory and $\mathcal{S}$ and $\mathcal{P}$ are real and pseudo-scalars respectively. The metric of a spherically symmetric charged AdS ModMax black hole is given by \begin{equation}\label{eq19}
ds^2=-f(r)^2dt^2+\frac{1}{f(r)^2}dr^2+r^2\left(d\theta^2+r^2\sin^2\theta d\phi^2\right)
\end{equation} 
with the metric function \begin{equation}\label{eq20}
f(r)=\frac{r^2}{l^2}-\frac{2 M}{r}+\frac{ q^2 e^{-\eta }}{r^2}+1
\end{equation}
The mass of the black hole can be obtained by setting $f(r_+)=0$, giving us
\begin{equation}\label{eq21}
M=\frac{\frac{r^4_+}{l^2}+e^{-\eta } q^2+r^2_+}{2 r_+}
\end{equation}
The temperature and entropy can also be calculated easily, which comes out to be
\begin{equation}\label{eq22}
T=\frac{\frac{3 r^4_+}{l^2}-e^{-\eta } q^2+r^2_+}{4 \pi  r^3_+}\quad \text{and}\quad S=\pi r^2_+
\end{equation}
Hence we can calculate the Gibbs free energy from the definition $F=M-TS$, which comes out to be \begin{equation}\label{eq23}
F=\frac{-\frac{r^4_+}{l^2}+3 e^{-\eta } q^2+r^2_+}{4 r_+}
\end{equation}
By employing dimensional analysis \cite{first}, we find that different physical properties scale with powers of $l$ enabling their representation in terms of appropriate scaling relations \begin{equation}\label{eq24}
\tilde{r}_+=r_+/l, \quad \tilde{q}=q/l , \quad \tilde{F}=F/l , \quad  \tilde{T}=T l , \quad \text{and}  \quad \tilde{M}=M/l.
\end{equation}
The critical points of different physical quantities can be determined by solving the corresponding set of equations \begin{equation}\label{eq25}
\frac{\partial\tilde{T}}{\partial \tilde{r}_+}=\frac{\partial^2\tilde{T}}{\partial^2 \tilde{r}_+}=0,
\end{equation}
Upon using the Hawking temperature and simultaneously solving the above equations we get the following set of critical values \begin{equation}\label{eq26}
\tilde{r}_{+c}=0.408248, \quad \tilde{q}_{c}=0.214004, \quad \tilde{T}_c=0.259899
\end{equation}where we have set $\eta=0.5$. And, by setting $q=0.5$ we get,\begin{equation}\label{eq27}
\tilde{r}_{+c}=0.408248, \quad \eta_{c}=2.19722, \quad \tilde{T}_c=0.259899
\end{equation}
\begin{figure}[h!]
    \centering
    \begin{subfigure}[b]{0.45\textwidth}
        \centering
        \includegraphics[width=\textwidth]{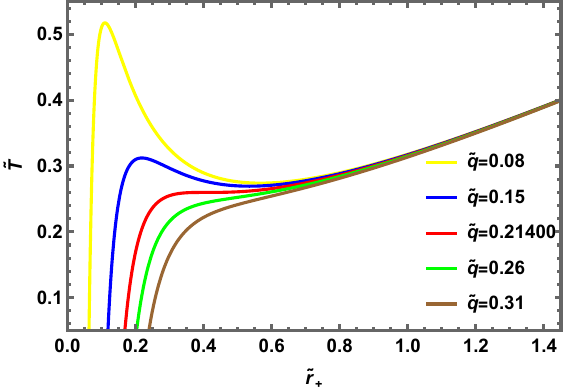}
        \caption{for $\eta=0.5$}
        \label{f1a}
    \end{subfigure}
    \hspace{0.8cm}
    \begin{subfigure}[b]{0.45\textwidth}
        \centering
        \includegraphics[width=\textwidth]{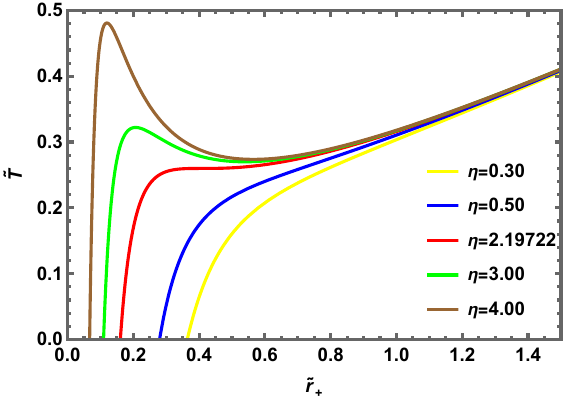}
        \caption{for $\tilde{q}=0.5$}
        \label{f1b}
    \end{subfigure}
     \caption{Hawking temperature as a function of horizon radius}
    \label{f1}
\end{figure}
In figure \ref{f1}, we show the variation of the Hawking temperature with the horizon radius $r_+$. Figure \ref{f1a} shows the variation of temperature for different values of $\tilde{q}$ and for a fixed value of ModMax parameter $\eta=0.5$. Here the yellow and the blue curve corresponds to $\tilde{q}<\tilde{q}_c$, the green and brown curve for $\tilde{q}>\tilde{q}_c$ and the red curve is for the critical value of charge. We observe that there are three black hole branches for $\tilde{q}$ values less than the critical $\tilde{q}$- the Small Black Hole (SBH) branch, the Intermediate Black Hole (IBH) branch and the Large Black Hole (LBH) branch. In figure \ref{f1b}, the variation of Hawking temperature for different values of $\eta$ is shown where we have fixed the value of $\tilde{q}$ at $0.5$. In this case we observe the three black hole branches appear for $\eta$ values greater the critical $\eta$ value- the brown and green curve as shown in figure \ref{f1b}. We show the three distinct branches for the fixed $\eta$ value of $0.5$ in the Gibbs free energy versus temperature plot in figure \ref{f2a}. Figure \ref{f2a} displays the familiar swallow-tail profile, which is a characteristic of a first-order phase transition.
In this case, three distinct black hole branches coexist for $\tilde{T}_b < \tilde{T} < \tilde{T}_a$, where $\tilde{T}_b$ and $\tilde{T}_a$ correspond to the temperatures at points $b$ and $a$, respectively. At point $p$, characterized by temperature $\tilde{T}_p$, a first-order phase transition occurs between the small and large black hole phases. Throughout this range, the free energy of the intermediate black hole (IBH) branch consistently exceeds that of both the small and large black hole phases, making the IBH state thermodynamically unfavorable and indicative of its instability in the overall phase diagram. From figure \ref{f2b} and \ref{f2c} we can observe that the swallow-tail behaviour exists for all $\tilde{q}<\tilde{q}_c$ and disappears for $\tilde{q}>\tilde{q}_c$ and the free energy becomes a smooth continuous curve.

\begin{figure}[h!]
    \centering
    \begin{subfigure}[b]{0.45\textwidth}
        \centering
        \includegraphics[width=\textwidth]{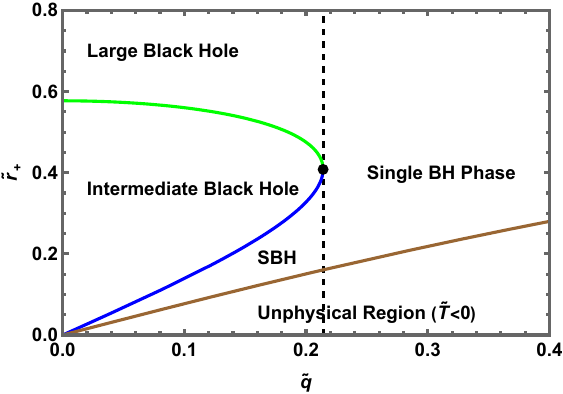}
        \caption{$\tilde{r}_+-\tilde{q}$ plane with fixed $\eta$}
        \label{fr1a}
    \end{subfigure}
    \hspace{0.8cm}
    \begin{subfigure}[b]{0.45\textwidth}
        \centering
        \includegraphics[width=\textwidth]{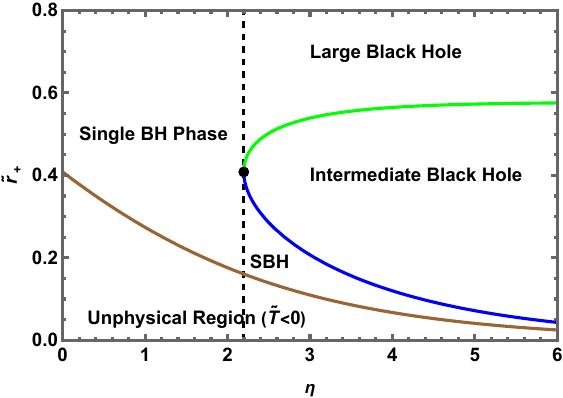}
        \caption{$\tilde{r}_+-\eta$ plane with fixed $\tilde{q}$}
        \label{fr1b}
    \end{subfigure}
     \caption{Phase structure in the parameter space}
    \label{fr1}
\end{figure}

In order to characterize the occurrence of different black hole phases in terms of the control parameters $\tilde{q}$ and $\eta$, we construct the parameter space diagram as shown in figure \ref{fr1}. In the left panel, the parameter space is drawn in the $\tilde{r_+}-\tilde{q}$ plane with fixed $\eta=0.5$. The area under the brown curve is the unphysical  region characterized by $\tilde{T}<0$. On the left side of the dashed line, the black hole phase structure splits into three distinct regions- the area below the blue curve represents the small black hole phase, the region between the blue and green curves corresponds to the intermediate black hole phase, and the region above the green curve denotes the large black hole phase. The black dot is precisely the critical point $(\tilde{q}_c, \tilde{r}_{+c})$ that marks the end of the first-order transition line. Towards the right of the dashed line, we have only one black hole phase. In the right panel, we have shown the parameter space in the $\tilde{r}_+-\eta$ plane with fixed $\tilde{q}=0.5$. Below the brown curve we have the unphysical region. The three black hole phases exist towards the right of the dashed line- small black hole phase under the blue curve, intermediate black hole phase between the blue and green curve and large black hole phase above the green curve. The black dot represents the critical point $(\eta_c, \tilde{r}_{+c})$ marking the end of the first-order transition line. A single black hole phase exists towards the left of the dashed line. The region where the single black hole and three distinct black hole phases exists is flipped in this parameter space because three black hole branches occur for $\eta>\eta_c$ (with fixed $\tilde{q}$) while the same happens for $\tilde{q}<\tilde{q}_c$ (with fixed $\eta$) as depicted in figure \ref{f1}

\section{Lyapunov exponents and phase transitions of ModMaX AdS black holes}\label{sec4}
In this section, we evaluate the Lyapunov exponents for both massive and massless particles in unstable circular orbits around the ModMax–AdS black hole background. We further demonstrate that the thermal behavior of the Lyapunov exponent can serve as a diagnostic tool, encoding signatures of the black hole’s phase transition. We begin our analysis with the case of massless particles.

\subsection{Maassless Particles}

\begin{figure}[h!]
    \centering
    \begin{subfigure}[b]{0.45\textwidth}
        \includegraphics[width=\textwidth]{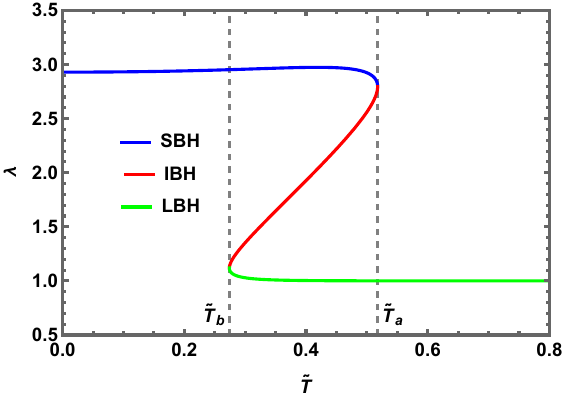}
        \caption{$\tilde{q}=0.08<\tilde{q}_c$}
        \label{f3a}
    \end{subfigure}
    \hspace{0.05\textwidth}
    \begin{subfigure}[b]{0.45\textwidth}
        \includegraphics[width=\textwidth]{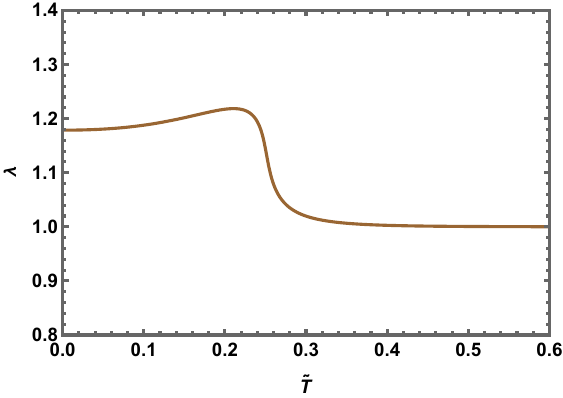}
        \caption{$\tilde{q}>\tilde{q}_c$}
        \label{f3b}
    \end{subfigure}
    
    \vspace{0.5cm} 

    \begin{subfigure}[b]{0.45\textwidth}
        \includegraphics[width=\textwidth]{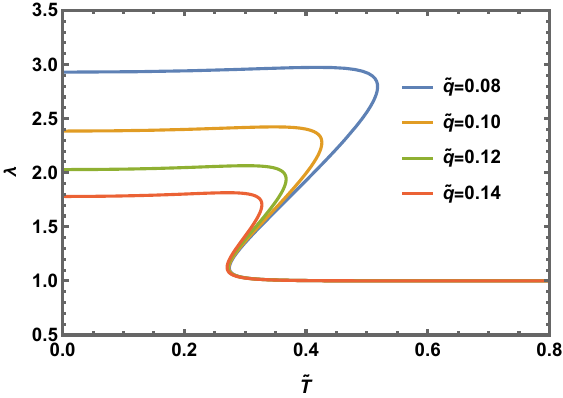}
        \caption{Lyapunov exponent versus temperature}
        \label{f3c}
    \end{subfigure}
    \hspace{0.05\textwidth}
    \begin{subfigure}[b]{0.45\textwidth}
        \includegraphics[width=\textwidth]{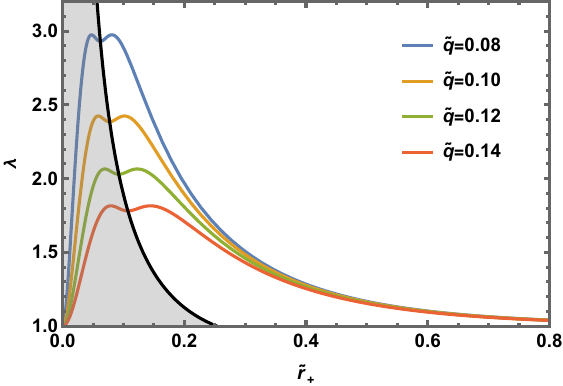}
        \caption{Lyapunov exponent versus horizon radius}
        \label{f3d}
    \end{subfigure}
    \caption{Lyapunov exponent $\lambda$ of massless particles as a function of temperature and horizon radius}
    \label{f3}
\end{figure}
The Lyapunov exponent for massless particles can be calculated using equations (\ref{eq5}) and (\ref{eq13}). The expression is cumbersome and not intuitive, so we just provide the plot of Lyapunov exponent versus temperature and horizon radius in figure \ref{f3}. The temperature dependence of the Lyapunov exponent is presented in figure \ref{f3a}, using the same color scheme as in figure \ref{f2}. For $\tilde{q}$ below the critical value $\tilde{q}_c$, the Lyapunov exponent exhibits a multivalued nature within the temperature interval $\tilde{T}_b$ to $\tilde{T}_a$. As the temperature rises, $\lambda$ decreases steadily until it reaches $\tilde{T}_b$, beyond which it becomes multivalued. This multivalued behaviour continues up to $\tilde{T}_a$. Within the range $\tilde{T}_b <\tilde{T}<\tilde{T}_a$, three distinct branches emerge: the blue and green curves correspond to the small and large black hole branches, respectively, while the red curve represents the intermediate black hole branch. This pattern of the Lyapunov exponent closely resembles the behaviour of the Gibbs free energy, indicating its potential as a diagnostic tool for identifying black hole phase transitions. For $\tilde{q}>\tilde{q}_c$, we saw the swallow-tail behaviour disappearing in figure \ref{f2c}. Consistently, we find that the multivalued behaviour of the Lyapunov exponent also vanishes for $\tilde{q} > \tilde{q}_c$, as illustrated in figure \ref{f3b}. We show the variation of $\lambda$ for differnt values of $\tilde{q}$ in figure \ref{f3c}. Here, it is evident that for higher temperatures, the Lyapunov exponent approaches a constant value across all values of $\tilde{q}$. To better understand the behaviour of the Lyapunov exponent, we present its variation with the horizon radius $r_+$ in figure \ref{f3d}. The grey-shaded region denotes the non-physical domain, where the Hawking temperature becomes negative, while the black curve corresponds to the case $\tilde{T}=0$. It is observed that the peak value of $\lambda$ increases as $\tilde{q}$ decreases. Moreover, irrespective of the value of $\tilde{q}$, the Lyapunov exponent approaches the same asymptotic value for sufficiently large $r_+$.

\subsection{Massive Particles}
\begin{figure}[h!]
    \centering
    \begin{subfigure}[b]{0.45\textwidth}
        \includegraphics[width=\textwidth]{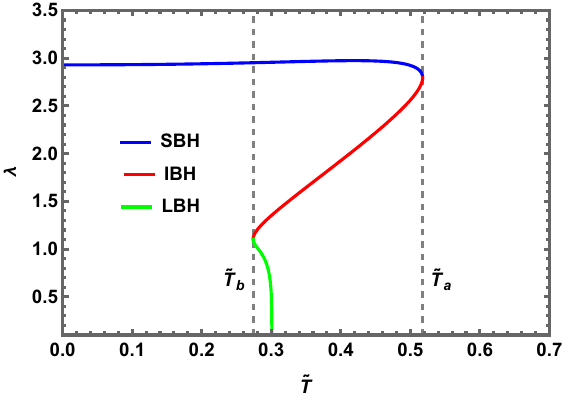}
        \caption{$\tilde{q}=0.08<\tilde{q}_c$}
        \label{f4a}
    \end{subfigure}
    \hspace{0.05\textwidth}
    \begin{subfigure}[b]{0.45\textwidth}
        \includegraphics[width=\textwidth]{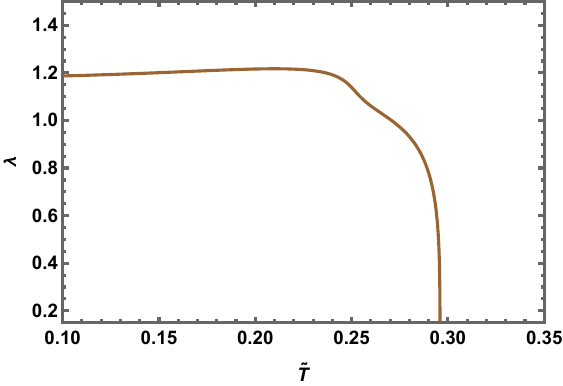}
        \caption{$\tilde{q}>\tilde{q}_c$}
        \label{f4b}
    \end{subfigure}
    
    \vspace{0.5cm} 

    \begin{subfigure}[b]{0.45\textwidth}
        \includegraphics[width=\textwidth]{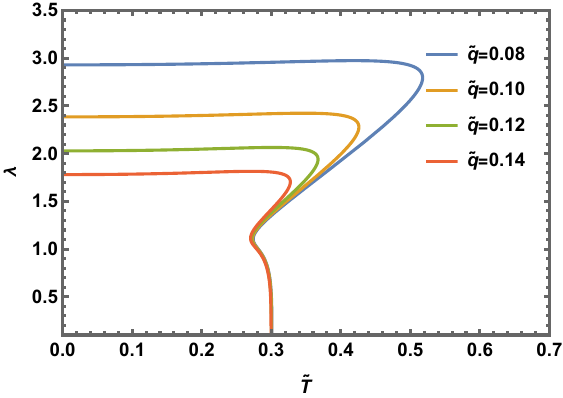}
        \caption{Lyapunov exponent versus temperature}
        \label{f4c}
    \end{subfigure}
    \hspace{0.05\textwidth}
    \begin{subfigure}[b]{0.45\textwidth}
        \includegraphics[width=\textwidth]{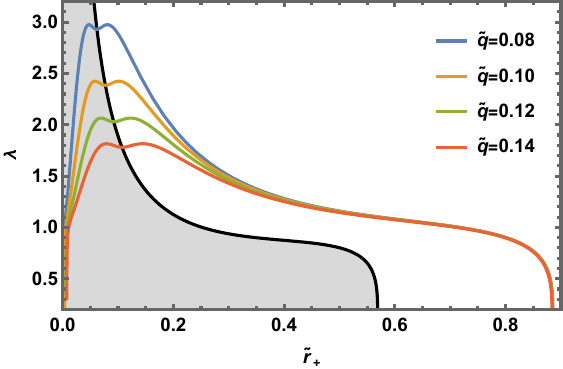}
        \caption{Lyapunov exponent versus horizon radius}
        \label{f4d}
    \end{subfigure}
    \caption{Lyapunov exponent $\lambda$ of massive particles as a function of temperature and horizon radius}
    \label{f4}
\end{figure}

We now shift our attention to computing Lyapunov exponent associated with massive particles. This can be done using equation (\ref{eq5}) and (\ref{eq16}). The variation of Lyapunov exponent with Hawking temperature and horizon radius is shown in figure \ref{f4}. The thermal profile of Lyapunov exponent is shown in figure \ref{f4a}. In the case of massive particles, both $r_0$ and $\lambda$ have dependency on the angular momentum $L$. In our analysis we have taken $L=20l$ as a representative choice for illustration. Similar to the massless case, the Lyapunov exponent exhibits a multivalued structure when the charge parameter lies below the critical threshold ($\tilde{q}<\tilde{q}_c$) within the temperature interval $\tilde{T}_b<\tilde{T}<\tilde{T}_a$ as seen in figure \ref{f4a}. Once the charge parameter exceeds this threshold, $\tilde{q}>\tilde{q}_c$ the multivalued nature disappears, and $\lambda$ becomes a smooth, single-valued function of temperature (figure \ref{f4b}). A notable difference from the massless scenario is that, for massive particles, the Lyapunov exponent tends to zero in the large black hole phase for all values of $\tilde{q}$, as illustrated in figure \ref{f4c}.  The variation of Lyapunov exponent $\lambda$ with the horizon radius is shown in figure \ref{f4d}, where the overall trend resembles that of the massless case: the maximum value of $\lambda$ increases as $\tilde{q}$ decreases. However, here $\lambda$ drops to zero at a particular horizon radius, which remains nearly the same for all $\tilde{q}$ values. This behaviour suggests that, in the large black hole regime, the dynamics of massive particles becomes non-chaotic, with $\lambda$ tending to $0$ signalling the absence of unstable equilibrium points.

We now provide a brief discussion on the physical mechanisms leading to the discontinuity of the Lyapunov exponent near the phase transition point. In general dynamical systems, the Lyapunov exponent depends on the linear stability matrix of the system around its typical trajectories (attractors). A phase transition of the system means that its macroscopic state (attractor) changes abruptly. For example in black hole, a transition from small to large horizon, in Ising model, a transition from ordered to disordered magnetization, in fluid dynamics, a transition from laminar to turbulent flow etc. Because the Lyapunov exponent depends on the stability of the trajectories around the attractor, any sudden change in attractor structure causes a sudden change in $\lambda$. In black hole backgrounds, the discontinuity of the Lyapunov exponent near a phase transition originates from the fact that unstable circular orbits are extremely sensitive to background geometry. The Lyapunov exponent is proportional to the square root of the second derivative of the effective potential evaluated at the unstable circular orbit radius (equation \ref{eq13} and \ref{eq16}) and the effective potential depends on the metric function. At the small/large black hole transition, the metric parameter (notably the horizon radius $r_+$ and near horizon curvature) changes abruptly between branches. Because $\lambda \sim\sqrt{V''_{\text{eff}}(r_0)}$, it inherits the abrupt change when the thermodynamically preferred branch switches.

\section{Critical exponents for ModMax AdS black holes with lyapunov exponent}\label{sec5}
In Landau’s theory of phase transitions, an order parameter is a quantity that distinguishes two phases by being zero in one and non-zero in the other, for example, magnetization in ferromagnets or density difference in liquid-gas systems.
Remarkably, it has been seen that in the small-large black hole transitions, this role can be played by the change in the Lyapunov exponent, serving as an effective order parameter. At the phase transition temperature $\tilde{T}_p$, the Lyapunov exponent for the small black hole branch is denoted by $\lambda_s$, while that for the large black hole branch is $\lambda_l$. At the second-order critical point, where $\tilde{T}_p = \tilde{T}_c$, these values coincide, $\lambda_s = \lambda_l = \lambda_c$, with $\lambda_c$ representing the Lyapunov exponent evaluated at the critical thermodynamic parameters. The difference in the Lyapunov exponent, $\Delta\lambda = \lambda_s - \lambda_l$, acts as an order parameter, vanishing at the critical point. To investigate the critical behaviour of $\lambda$, we determine the critical exponent - a parameter describing how a physical quantity scales near a phase transition’s critical point. For this analysis, we follow the approach described in \cite{le2,ce} and compute the exponent $\delta$, which satisfies
\begin{equation}\label{eq37}
\Delta\lambda\equiv\lambda_s-\lambda_l=\mid\tilde{T}-\tilde{T}_c\mid^\delta
\end{equation}
We begin by writing the expressions for the horizon radius and the Hawking temperature at the phase transition point as
 \begin{equation}\label{eq38}
\tilde{r}_p=\tilde{r}_c\left(1+\Delta\right) \quad \text{and} \quad \tilde{T}(\tilde{r}_+)=\tilde{T}_c\left(1+\epsilon\right)
\end{equation}
where $\mid\Delta\mid\ll1$ and $\mid\epsilon\mid\ll1$. We then perform a Taylor expansion of the Lyapunov exponent about the critical point $\tilde{r}_c$
\begin{equation}\label{eq39}
\lambda=\lambda_c+\left[\frac{\partial \lambda}{\partial \tilde{r}_+}\right]d\tilde{r}_++\mathcal{O}\left(\tilde{r}_+\right)
\end{equation}
Here the subscript $c$ denotes quantities evaluated at the critical point. Using Eqs. (\ref{eq38}) and (\ref{eq39}), we then obtain,
\begin{equation}\label{eq40}
\frac{\Delta\lambda}{\lambda_c}=\frac{\lambda_s-\lambda_l}{\lambda_c}=\frac{\tilde{r}_c}{\lambda}_c\left[\frac{\partial\lambda}{\partial\tilde{r}_+}\right]_c\left(\Delta_s-\Delta_l\right)
\end{equation}
Here, we also make use of the fact that, at the critical point,t $\lambda_s = \lambda_l = \lambda_c$ and so $\lambda_s(\tilde{r}_c)-\lambda_l(\tilde{r}_c)=0$.
Similarly, we perform a Taylor expansion of the Hawking temperature around the critical horizon radius $\tilde{r}_c$ and obtain 
\begin{equation}\label{eq41}
\tilde{T}=\tilde{T}_c+\frac{\tilde{r}^2_c}{2}\left[\frac{\partial^2 \tilde{T}}{\partial \tilde{r}^2_+}\right]\Delta^2
\end{equation}
Here, higher-order terms are neglected, and the condition $\left[\frac{\partial\tilde{T}}{\partial\tilde{r}_+}\right]_c \to 0$ is applied. Using Eqs. (\ref{eq40}) and (\ref{eq41}), we then arrive at a simplified form,
\begin{equation}\label{eq42}
\frac{\Delta\lambda}{\lambda_c}=k\sqrt{t-1}
\end{equation}
where $t=\frac{\tilde{T}}{\tilde{T}_c}$ and 
\begin{equation}\label{eq43}
k=\frac{\sqrt{\tilde{T}_c}}{\lambda_c}\left[\frac{\partial\Delta\lambda}{\partial\tilde{r}_+}\right]_c\left[\frac{1}{2}\frac{\partial^2\tilde{T}}{\partial\tilde{r}^2_+}\right]^{-1/2}_c
\end{equation}
Thus, it follows that the critical exponent $\delta$ associated with the order parameter $\Delta\lambda$ takes the value $1/2$ in the vicinity of the critical point.

\subsection{Numerical Verification}
We next provide a numerical confirmation of the previously discussed result for ModMax AdS black holes, considering both massive and massless particle cases. 
\begin{figure}[h!]
    \centering
    \begin{subfigure}[b]{0.45\textwidth}
        \centering
        \includegraphics[width=\textwidth]{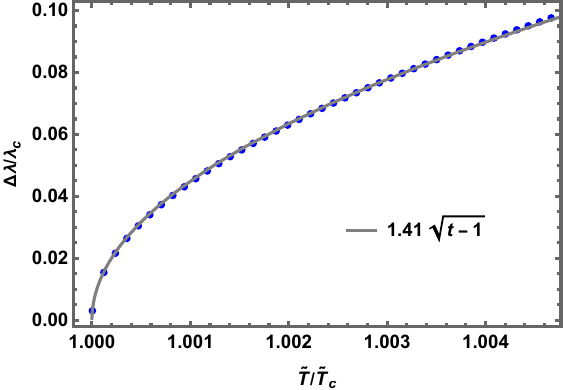}
        \caption{for massless particles}
        \label{f5a}
    \end{subfigure}
    \hspace{0.8cm}
    \begin{subfigure}[b]{0.45\textwidth}
        \centering
        \includegraphics[width=\textwidth]{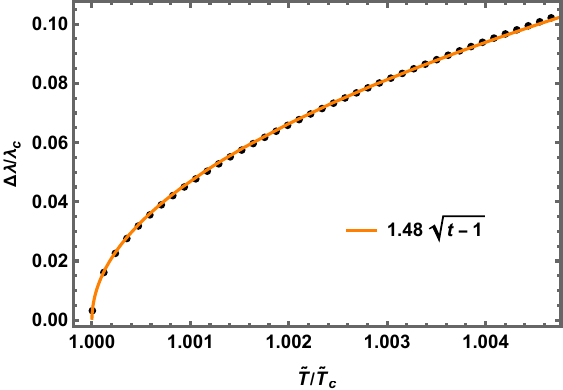}
        \caption{for massive particles}
        \label{f5b}
    \end{subfigure}
     \caption{Rescaled discontinuity in Lyapunov exponent,$\Delta\lambda/\lambda_c$ versus rescaled phase transition temperature $t$}
    \label{f5}
\end{figure}
The rescaled order parameter, $\Delta\lambda/\lambda_c$, is plotted against the reduced phase transition temperature, $t = \tilde{T}_p / \tilde{T}_c$. For the massless case, figure \ref{f5a} depicts the behaviour of $\Delta\lambda/\lambda_c$ as a function of $t$, focusing on the region close to the critical value. In the figure, the discrete points correspond to the numerically obtained data, while the smooth, coloured curves represent the fitted results. We similarly plot $\Delta\lambda/\lambda_c$ versus $\tilde{T}_p / \tilde{T}_c$ for the massive particle case as shown in figure \ref{f5b}. Our analysis shows that the relation $\Delta\lambda/\lambda_c = k\sqrt{t - 1}$ provides an excellent fit to the numerically obtained data for both massless and massive particles with $k=1.41963$ and $1.48407$ respectively. While the constant $k$ takes different values in the two cases, the critical exponent $\delta$ remains unchanged at $1/2$.

\section{Violation of Chaos Bound}\label{sec6}
In black hole physics, the Lyapunov exponent is subject to a fundamental upper restriction, commonly referred to as the chaos bound \cite{mss}. This limit, $\lambda \le \frac{2\pi T}{\hbar}$, signifies the fastest possible growth rate of chaos in thermal quantum systems. Maldacena, Shenker, and Stanford established this result through arguments in quantum field theory and analyses involving shock waves near event horizons. Later, Hashimoto and Tanahashi extended the framework to describe the motion of test particles close to the horizon, arriving at the condition $\lambda \le \kappa$, where $\kappa$ denotes the surface gravity \cite{Hashimoto}. Since $\kappa = 2\pi T$ by thermodynamic identity, this unifies the particle dynamics perspective with the quantum field theory formulation of the chaos bound. Their work thus revealed a deep correspondence between the dynamical instability of trajectories near black holes and the universal limit on chaotic behaviour. In this section we investigate the violation of the chaos bound by taking the difference of $\lambda$ and $\kappa$. Given that the bound dictates $\lambda \le \kappa$, a positive value of $\lambda - \kappa$ indicates a violation, whereas a non-positive value implies that the bound is preserved.

\begin{figure}[h!]
    \centering
    \begin{subfigure}[b]{0.45\textwidth}
        \centering
        \includegraphics[width=\textwidth]{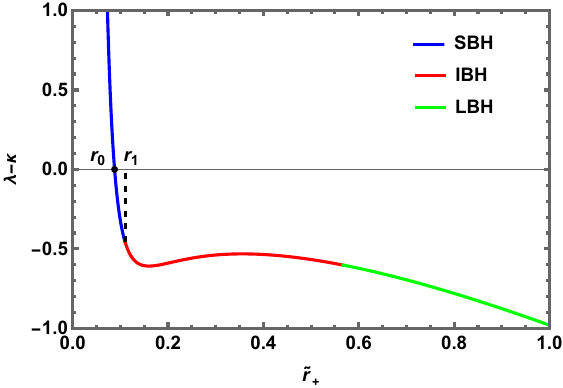}
        \caption{for fixed $\tilde{q}=0.08$ and $\eta=0.5$}
        \label{f6a}
    \end{subfigure}
    \hspace{0.8cm}
    \begin{subfigure}[b]{0.45\textwidth}
        \centering
        \includegraphics[width=\textwidth]{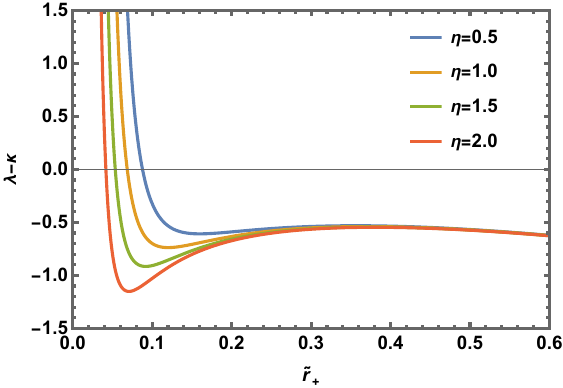}
        \caption{for different values of $\eta$}
        \label{f6b}
    \end{subfigure}
     \caption{Variation of $\lambda - \kappa$ with the horizon radius $\tilde{r}_+$ indicating regions where the chaos bound may be violated in case of massless particles}
    \label{f6}
\end{figure}
Figure \ref{f6a} shows the difference between Lyapunov exponent and surface gravity as a function of horizon radius for fixed values of $\tilde{q}$ and $\eta$. We can observe from this plot that there is a violation of the chaos bound when the horizon radius $r_+$ is less than $r_0$ marked by the black dot in figure \ref{f6a}. It is also observed that the bound is violated in the small, thermodynamically stable black hole phase represented by the blue curve in the figure while in the non-small black hole phases, the bound remains preserved. Notably, the authors in \cite{Lei} has shown that the chaos bound can only be violated within the thermodynamically stable phases of the black hole. Our findings support this claim, as the region of bound violation ($r_0=0.0878734$) lies entirely within the stable phase, extending up to $r_1=0.109925$, where the unstable intermediate black hole phase begins. The stability region can be easily verified by checking whether the specific heat is positive. We also present in figure \ref{f6b} the variation of $\lambda - \kappa$ with the horizon radius $\tilde{r}_+$ for different values of the ModMax parameter $\eta$, keeping $\tilde{q}$ fixed. It is observed that as $\eta$ increases, the point $r_0$ at which $\lambda - \kappa$ becomes positive, indicating the onset of chaos bound violation, shifts to smaller values of $\tilde{r}_+$. Consequently, the range of the horizon radius over which the chaos bound is violated becomes progressively narrower for larger $\eta$. The reason for this shift could be the following- ModMax parameter $\eta$ enhances nonlinear electromagnetic effects in the background. This modification alters the effective potential (and hence the unstable circular orbits) felt by test particles, and also modifies the surface gravity $\kappa$ As $\eta$ increases, these combined changes make $\lambda$ exceed $\kappa$ at smaller radii, producing the observed inward shift of $r_0$.
\begin{figure}[h!]
    \centering
    \begin{subfigure}[b]{0.45\textwidth}
        \centering
        \includegraphics[width=\textwidth]{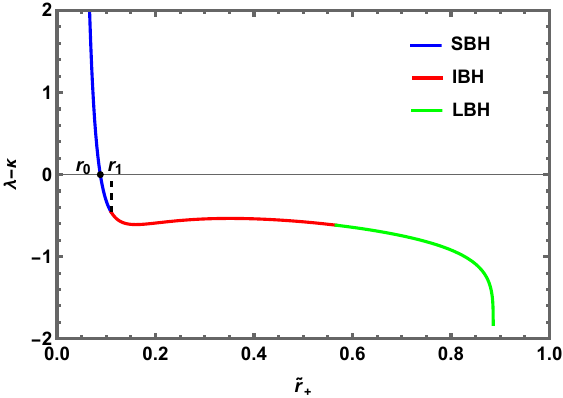}
        \caption{for fixed $\tilde{q}=0.08$, $\eta=0.5$ and $L=20l$}
        \label{f7a}
    \end{subfigure}
    \hspace{0.8cm}
    \begin{subfigure}[b]{0.45\textwidth}
        \centering
        \includegraphics[width=\textwidth]{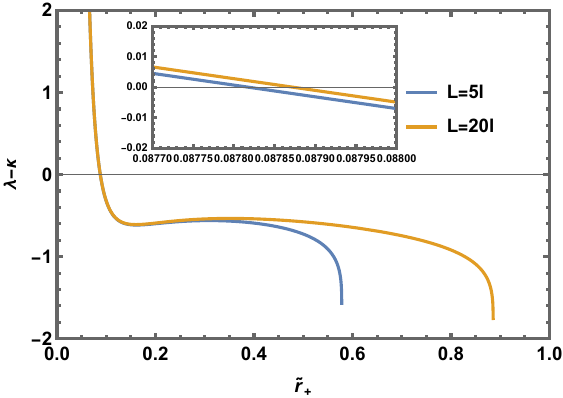}
        \caption{for different values of angular momentum}
        \label{f7b}
    \end{subfigure}
     \caption{Variation of $\lambda - \kappa$ with the horizon radius $\tilde{r}_+$ indicating regions where the chaos bound may be violated in case of massive particles}
    \label{f7}
\end{figure}
In figure \ref{f7} we show the variation of $\lambda-\kappa$ with respect to the horizon radius for massive particles, where the Lyapunov exponent depends on the angular momentum $L$. In figure \ref{f7a} we again observe just like the massless particle case that the chaos bound is violated in the thermodynamically stable, small black hole branch where $\lambda-\kappa>0$. In figure \ref{f7b} we show the same curve for two different values of the angular momentum $L$. From the figure we can observe that increasing the angular momentum $L$, increases the corresponding horizon radius $r_0$ at which the chaos bound is violated. The same result was also observed in \cite{Lei} and recently in \cite{le9}. The above cases were for $\tilde{q}<\tilde{q}_c$ where a phase transition occurs. We now examine the case $\tilde{q}>\tilde{q}_c$ where no phase transition takes place.
\begin{figure}[h!]
    \centering
    \begin{subfigure}[b]{0.45\textwidth}
        \centering
        \includegraphics[width=\textwidth]{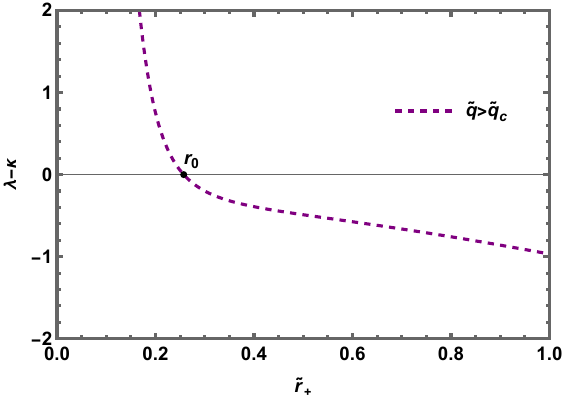}
        \caption{for massless particles with $\tilde{q}=0.25>\tilde{q}_c$}
        \label{f8a}
    \end{subfigure}
    \hspace{0.8cm}
    \begin{subfigure}[b]{0.45\textwidth}
        \centering
        \includegraphics[width=\textwidth]{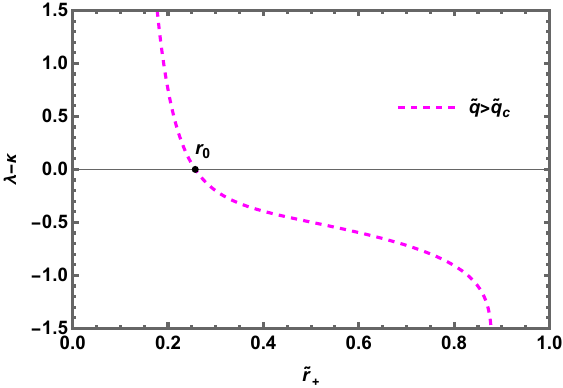}
        \caption{for massive particles with $\tilde{q}=0.25>\tilde{q}_c$}
        \label{f8b}
    \end{subfigure}
     \caption{Variation of $\lambda - \kappa$ with the horizon radius $\tilde{r}_+$ indicating regions where the chaos bound may be violated when there is no phase transition ($\tilde{q}>\tilde{q}_c$)}
    \label{f8}
\end{figure}
From figures \ref{f8a} and \ref{f8b}, we observe that, despite the absence of a phase transition, the chaos bound is still violated in both the massless and massive particle cases when the horizon radius falls below the threshold value $r_0$

We may pause here to dive deeper into the physical implications of the chaos bound and its violation. We revisit the original idea of Maldacena, Shenker and Stanford where they showed that in a system having large number of degrees of freedom with a semiclassical gravity dual, the growth of chaos, which they quantified through the exponential growth of out-of-time-ordered correlators, is universally bounded. In their framework, the Lyapunov exponent quantified how quickly small perturbations grow in the system, while the scrambling time ($t_*$) specified the duration needed for information to become fully dispersed across the microscopic degrees of freedom (so a larger $\lambda$ would mean faster scrambling). However we have seen this bound to be violated in many cases including our own system. Physically, this may indicate that the information scrambling in the thermodynamically stable small black hole phase support faster scrambling dynamics and shorter scrambling times than those captured by semiclassical holographic arguments. Moreover, the fact that the violation persists even in the absence of a phase transition ($\tilde{q}>\tilde{q}_c$), hints at a more general mechanism tied to the nonlinear structure of the theory rather than critical phenomena alone.

\section{conclusion and discussion}\label{sec7}
In this work, we have explored the connection between Lyapunov exponents and phase transitions of ModMax AdS black holes. We computed the Lyapunov exponent associated with massless and massive particles in circular orbit around the black hole and show its variation with the Hawking temperature. Our analysis reveals that, for $\tilde{q}$ values below the critical threshold $\tilde{q}_c$, the Lyapunov exponent as a function of temperature develops a multivalued structure. These distinct branches correspond to the small black hole (SBH), intermediate black hole (IBH), and large black hole (LBH) phases. When $\tilde{q}$ exceeds $\tilde{q}_c$, this multivalued nature vanishes, mirroring the qualitative behaviour of the free energy. This pattern emerges for both massless and massive particle orbits, indicating that the Lyapunov exponent encodes signatures of the black hole’s thermodynamic phase structure.

We further examined the jump in the Lyapunov exponent, $\Delta\lambda$, for both massless and massive probes, and demonstrated that it can serve as an effective order parameter characterizing the black hole phase transition. From an analytical standpoint, we established that the associated critical exponent $\delta$ takes the exact value $1/2$ which is identical to that of the van der Waals fluid in the mean-field framework. We verify this theoretical result through numerical calculations of $\Delta\lambda/\lambda_c$ near the critical point, finding excellent agreement with the predicted scaling. Hence, our analysis further reinforces the universality of the critical exponent, consistently obtaining the value $1/2$.

We further investigated the violation of chaos bound in ModMax AdS black hole background and found that the bound is observed to be violated in the thermodynamically stable, small black hole branch while it is remains preserved in the non-small black hole branch. The violation sets in when the horizon radius falls below a threshold value, denoted by $r_0$, irrespective of whether the black hole undergoes a phase transition. We find that increasing the ModMax parameter lowers the threshold value. We suspect it is because increasing the ModMax parameter $\eta$, strengthens nonlinear electromagnetic effects, altering the effective potential and surface gravity so that $\lambda$ exceeds $\kappa$ at smaller radii. We also observed that in the massive particle case, increasing the angular momentum $L$ increases the corresponding $r_0$ at which chaos bound is violated.  Hence, our analysis points toward a compelling link between the thermal stability of the black hole and the violation of the chaos bound. This suggests that the stability properties of the black hole may play a decisive role in determining whether the bound is preserved or breached. Uncovering the precise nature of this connection could provide valuable insight into the interplay between black hole thermodynamics, nonlinear electrodynamics, and quantum chaotic behaviour. Moreover, the precise physical process driving this violation, the underlying microscopic degrees of freedom responsible for it, and its broader implications remain elusive. Addressing these questions is essential for revealing the deeper principles that unify black hole thermodynamics and the dynamics of chaos, two domains that appear distinct yet are profoundly interlinked.


\begin{thebibliography}{99}

\bibitem{Phys}
S.~W.~Hawking,
``Gravitational radiation from colliding black holes,''
Phys. Rev. Lett. \textbf{26}, 1344-1346 (1971)
doi:10.1103/PhysRevLett.26.1344
\bibitem{bekens}
J.~Bekenstein,
``Bekenstein-Hawking entropy,''
Scholarpedia \textbf{3}, no.10, 7375 (2008)
doi:10.4249/scholarpedia.7375
\bibitem{Hawking}
S.~W.~Hawking,
``Black hole explosions,''
Nature \textbf{248}, 30-31 (1974)
doi:10.1038/248030a0
\bibitem{Hawking2}
S.~W.~Hawking,
``Particle Creation by Black Holes,''
Commun. Math. Phys. \textbf{43}, 199-220 (1975)
[erratum: Commun. Math. Phys. \textbf{46}, 206 (1976)]
doi:10.1007/BF02345020
\bibitem{Bardeen}
J.~M.~Bardeen, B.~Carter and S.~W.~Hawking,
``The Four laws of black hole mechanics,''
Commun. Math. Phys. \textbf{31}, 161-170 (1973)
doi:10.1007/BF01645742

\bibitem{Davies}
P.~C.~W.~Davies,
``Thermodynamics of Black Holes,''
Proc. Roy. Soc. Lond. A \textbf{353}, 499-521 (1977)
doi:10.1098/rspa.1977.0047
\bibitem{Hut}
P. Hut,``Charged black holes and phase transitions,”
Monthly Notices of the Royal
 Astronomical Society, vol. 180, pp. 379–389, 10 1977
 

\bibitem{Maldacena}
J.~M.~Maldacena,
``The Large N limit of superconformal field theories and supergravity,''
Adv. Theor. Math. Phys. \textbf{2}, 231-252 (1998)
doi:10.4310/ATMP.1998.v2.n2.a1
[arXiv:hep-th/9711200 [hep-th]].

\bibitem{Kubiz}
D.~Kubiznak and R.~B.~Mann,
``P-V criticality of charged AdS black holes,''
JHEP \textbf{07}, 033 (2012)
doi:10.1007/JHEP07(2012)033
[arXiv:1205.0559 [hep-th]].

\bibitem{Hawkpage}
S.~W.~Hawking and D.~N.~Page,
``Thermodynamics of Black Holes in anti-De Sitter Space,''
Commun. Math. Phys. \textbf{87}, 577 (1983)
doi:10.1007/BF01208266

\bibitem{Cai}
R.~G.~Cai, L.~M.~Cao, L.~Li and R.~Q.~Yang,
``P-V criticality in the extended phase space of Gauss-Bonnet black holes in AdS space,''
JHEP \textbf{09}, 005 (2013)
doi:10.1007/JHEP09(2013)005
[arXiv:1306.6233 [gr-qc]].

\bibitem{Kastor}
D.~Kastor, S.~Ray and J.~Traschen,
``Enthalpy and the Mechanics of AdS Black Holes,''
Class. Quant. Grav. \textbf{26}, 195011 (2009)
doi:10.1088/0264-9381/26/19/195011
[arXiv:0904.2765 [hep-th]].

\bibitem{Dolan}
B.~P.~Dolan,
``The cosmological constant and the black hole equation of state,''
Class. Quant. Grav. \textbf{28}, 125020 (2011)
doi:10.1088/0264-9381/28/12/125020
[arXiv:1008.5023 [gr-qc]].

\bibitem{Dolan2}
B.~P.~Dolan,
``Pressure and volume in the first law of black hole thermodynamics,''
Class. Quant. Grav. \textbf{28}, 235017 (2011)
doi:10.1088/0264-9381/28/23/235017
[arXiv:1106.6260 [gr-qc]].

\bibitem{Dolan3}
B.~P.~Dolan,
``Compressibility of rotating black holes,''
Phys. Rev. D \textbf{84}, 127503 (2011)
doi:10.1103/PhysRevD.84.127503
[arXiv:1109.0198 [gr-qc]].


\bibitem{Kubizna}
D.~Kubiznak, R.~B.~Mann and M.~Teo,
``Black hole chemistry: thermodynamics with Lambda,''
Class. Quant. Grav. \textbf{34}, no.6, 063001 (2017)
doi:10.1088/1361-6382/aa5c69
[arXiv:1608.06147 [hep-th]].

\bibitem{Xu}
W.~Xu, H.~Xu and L.~Zhao,
``Gauss-Bonnet coupling constant as a free thermodynamical variable and the associated criticality,''
Eur. Phys. J. C \textbf{74}, 2970 (2014)
doi:10.1140/epjc/s10052-014-2970-8
[arXiv:1311.3053 [gr-qc]].

\bibitem{Xu2}
W.~Xu and L.~Zhao,
``Critical phenomena of static charged AdS black holes in conformal gravity,''
Phys. Lett. B \textbf{736}, 214-220 (2014)
doi:10.1016/j.physletb.2014.07.019
[arXiv:1405.7665 [gr-qc]].

\bibitem{Zhang}
M.~Zhang, D.~C.~Zou and R.~H.~Yue,
``Reentrant phase transitions and triple points of topological AdS black holes in Born-Infeld-massive gravity,''
Adv. High Energy Phys. \textbf{2017}, 3819246 (2017)
doi:10.1155/2017/3819246
[arXiv:1707.04101 [hep-th]].


\bibitem{Ruppeiner:2012uc}
G.~Ruppeiner,
Thermodynamic curvature: pure fluids to black holes,
J. Phys. Conf. Ser. \textbf{410}, 012138 (2013)
doi:10.1088/1742-6596/410/1/012138
[arXiv:1210.2011 [gr-qc]].
\bibitem{Miao:2017cyt}
Y.~G.~Miao and Z.~M.~Xu,
Microscopic structures and thermal stability of black holes conformally coupled to scalar fields in five dimensions,
Nucl. Phys. B \textbf{942}, 205-220 (2019)
doi:10.1016/j.nuclphysb.2019.03.015
[arXiv:1711.01757 [hep-th]].
\bibitem{Guo:2019oad}
X.~Y.~Guo, H.~F.~Li, L.~C.~Zhang and R.~Zhao,
Microstructure and continuous phase transition of a Reissner-Nordstrom-AdS black hole,
Phys. Rev. D \textbf{100}, no.6, 064036 (2019)
doi:10.1103/PhysRevD.100.064036
[arXiv:1901.04703 [gr-qc]].
\bibitem{Wei:2019yvs}
S.~W.~Wei, Y.~X.~Liu and R.~B.~Mann,
Ruppeiner Geometry, Phase Transitions, and the Microstructure of Charged AdS Black Holes,
Phys. Rev. D \textbf{100}, no.12, 124033 (2019)
doi:10.1103/PhysRevD.100.124033
[arXiv:1909.03887 [gr-qc]].
\bibitem{Wang:2019cax}
P.~Wang, H.~Wu and H.~Yang,
Thermodynamic Geometry of AdS Black Holes and Black Holes in a Cavity,
Eur. Phys. J. C \textbf{80}, no.3, 216 (2020)
doi:10.1140/epjc/s10052-020-7776-2
[arXiv:1910.07874 [gr-qc]].
\bibitem{Yerra:2020oph}
P.~K.~Yerra and C.~Bhamidipati,
Ruppeiner Geometry, Phase Transitions and Microstructures of Black Holes in Massive Gravity,
Int. J. Mod. Phys. A \textbf{35}, no.22, 2050120 (2020)
doi:10.1142/S0217751X20501201
[arXiv:2006.07775 [hep-th]].
\bibitem{Yerra:2021hnh}
P.~K.~Yerra and C.~Bhamidipati,
Novel relations in massive gravity at Hawking-Page transition,
Phys. Rev. D \textbf{104}, no.10, 104049 (2021)
doi:10.1103/PhysRevD.104.104049
[arXiv:2107.04504 [gr-qc]].


\bibitem{Wu:2022whe}
D.~Wu,
Topological classes of rotating black holes,
Phys. Rev. D \textbf{107}, no.2, 024024 (2023)
doi:10.1103/PhysRevD.107.024024
[arXiv:2211.15151 [gr-qc]].
\bibitem{Liu:2022aqt}
C.~Liu and J.~Wang,
Topological natures of the Gauss-Bonnet black hole in AdS space,
Phys. Rev. D \textbf{107}, no.6, 064023 (2023)
doi:10.1103/PhysRevD.107.064023
[arXiv:2211.05524 [gr-qc]].
\bibitem{Fan:2022bsq}
Z.~Y.~Fan,
Topological interpretation for phase transitions of black holes,
Phys. Rev. D \textbf{107}, no.4, 044026 (2023)
doi:10.1103/PhysRevD.107.044026
[arXiv:2211.12957 [gr-qc]].

\bibitem{Gogoi:2023xzy}
N.~J.~Gogoi and P.~Phukon,
Thermodynamic topology of 4D dyonic AdS black holes in different ensembles,
Phys. Rev. D \textbf{108}, no.6, 066016 (2023)
doi:10.1103/PhysRevD.108.066016
[arXiv:2304.05695 [hep-th]].



\bibitem{Ali:2023zww}
M.~S.~Ali, H.~El Moumni, J.~Khalloufi and K.~Masmar,
Topology of Born-Infeld-AdS Black Hole Phase Transition,
[arXiv:2306.11212 [hep-th]].

\bibitem{Saleem:2023oue}
M.~A.~Saleem and A.~Taani,
The chaotic behavior of black holes: Investigating a topological retraction in anti-de Sitter spaces,
New Astron. \textbf{107}, 102149 (2024)
doi:10.1016/j.newast.2023.102149
\bibitem{Shahzad:2023cis}
M.~U.~Shahzad, A.~Mehmood, S.~Sharif and A.~\"Ovg\"un,
Criticality and topological classes of neutral Gauss\textendash{}Bonnet AdS black holes in 5D,
Annals Phys. \textbf{458}, no.3, 169486 (2023)
doi:10.1016/j.aop.2023.169486
\bibitem{Chen:2023elp}
Z.~Q.~Chen and S.~W.~Wei,
Thermodynamics, Ruppeiner geometry, and topology of Born-Infeld black hole in asymptotic flat spacetime,
Nucl. Phys. B \textbf{996}, 116369 (2023)
doi:10.1016/j.nuclphysb.2023.116369
\bibitem{Bai:2022klw}
N.~C.~Bai, L.~Li and J.~Tao,
Topology of black hole thermodynamics in Lovelock gravity,
Phys. Rev. D \textbf{107}, no.6, 064015 (2023)
doi:10.1103/PhysRevD.107.064015
[arXiv:2208.10177 [gr-qc]].
\bibitem{Yerra:2022alz}
P.~K.~Yerra and C.~Bhamidipati,
Topology of black hole thermodynamics in Gauss-Bonnet gravity,
Phys. Rev. D \textbf{105}, no.10, 104053 (2022)
doi:10.1103/PhysRevD.105.104053
[arXiv:2202.10288 [gr-qc]].
\bibitem{Hazarika:2023iwp}
B.~Hazarika and P.~Phukon,
Thermodynamic Topology of $D=4,5$ Horava Lifshitz Black Hole in Two Ensembles,
[arXiv:2312.06324 [hep-th]].


\bibitem{Liu:2014gvf}
Y.~Liu, D.~C.~Zou and B.~Wang,
Signature of the Van der Waals like small-large charged AdS black hole phase transition in quasinormal modes,
JHEP \textbf{09}, 179 (2014)
doi:10.1007/JHEP09(2014)179
[arXiv:1405.2644 [hep-th]].
\bibitem{Zou:2017juz}
D.~C.~Zou, Y.~Liu and R.~H.~Yue,
Behavior of quasinormal modes and Van der Waals-like phase transition of charged AdS black holes in massive gravity,
Eur. Phys. J. C \textbf{77}, no.6, 365 (2017)
doi:10.1140/epjc/s10052-017-4937-z
[arXiv:1702.08118 [gr-qc]].
\bibitem{Zhang:2020khz}
M.~Zhang, C.~M.~Zhang, D.~C.~Zou and R.~H.~Yue,
Phase transition and Quasinormal modes for Charged black holes in 4D Einstein-Gauss-Bonnet gravity,
Chin. Phys. C \textbf{45}, no.4, 045105 (2021)
doi:10.1088/1674-1137/abe19a
[arXiv:2009.03096 [hep-th]].
\bibitem{Mahapatra:2016dae}
S.~Mahapatra,
Thermodynamics, Phase Transition and Quasinormal modes with Weyl corrections,
JHEP \textbf{04}, 142 (2016)
doi:10.1007/JHEP04(2016)142
[arXiv:1602.03007 [hep-th]].
\bibitem{Chabab:2016cem}
M.~Chabab, H.~El Moumni, S.~Iraoui and K.~Masmar,
Behavior of quasinormal modes and high dimension RN\textendash{}AdS black hole phase transition,
Eur. Phys. J. C \textbf{76}, no.12, 676 (2016)
doi:10.1140/epjc/s10052-016-4518-6
[arXiv:1606.08524 [hep-th]].


\bibitem{Wei:2017mwc}
S.~W.~Wei and Y.~X.~Liu,
Photon orbits and thermodynamic phase transition of $d$-dimensional charged AdS black holes,
Phys. Rev. D \textbf{97}, no.10, 104027 (2018)
doi:10.1103/PhysRevD.97.104027
[arXiv:1711.01522 [gr-qc]].
\bibitem{Wei:2018aqm}
S.~W.~Wei, Y.~X.~Liu and Y.~Q.~Wang,
Probing the relationship between the null geodesics and thermodynamic phase transition for rotating Kerr-AdS black holes,
Phys. Rev. D \textbf{99}, no.4, 044013 (2019)
doi:10.1103/PhysRevD.99.044013
[arXiv:1807.03455 [gr-qc]].
\bibitem{Zhang:2019tzi}
M.~Zhang, S.~Z.~Han, J.~Jiang and W.~B.~Liu,
Phys. Rev. D \textbf{99}, no.6, 065016 (2019)
doi:10.1103/PhysRevD.99.065016
[arXiv:1903.08293 [hep-th]].



\bibitem{Zhang:2019glo}
M.~Zhang and M.~Guo,
Can shadows reflect phase structures of black holes?,
Eur. Phys. J. C \textbf{80}, no.8, 790 (2020)
doi:10.1140/epjc/s10052-020-8389-5
[arXiv:1909.07033 [gr-qc]].
\bibitem{Belhaj:2020nqy}
A.~Belhaj, L.~Chakhchi, H.~El Moumni, J.~Khalloufi and K.~Masmar,
Thermal Image and Phase Transitions of Charged AdS Black Holes using Shadow Analysis,
Int. J. Mod. Phys. A \textbf{35}, no.27, 2050170 (2020)
doi:10.1142/S0217751X20501705
[arXiv:2005.05893 [gr-qc]].

\bibitem{lyp}
A.M Lyapunov,
The general problem of the stability of motion
Int. J. Control 55, 531 (1992).
doi:https://doi.org/10.1080/00207179208934253
\bibitem{lyp2}
M. Sandri,
Numerical calculation of lyapunov exponents,
Math. J. 6, 78 (1996).


\bibitem{syk}
N.~Sorokhaibam,
``Phase transition and chaos in charged SYK model,''
JHEP \textbf{07}, 055 (2020)
doi:10.1007/JHEP07(2020)055
[arXiv:1912.04326 [hep-th]].
\bibitem{syk2}
A.~Davis and Y.~Wang,
``Quantum chaos and phase transition in the Yukawa\textendash{}Sachdev-Ye-Kitaev model,''
Phys. Rev. B \textbf{107}, no.20, 205122 (2023)
doi:10.1103/PhysRevB.107.205122
[arXiv:2212.03265 [cond-mat.str-el]].
\bibitem{Dicke}
C.~Emary and T.~Brandes,
``Chaos and the quantum phase transition in the Dicke model,''
Phys. Rev. E \textbf{67}, 066203 (2003)
doi:10.1103/PhysRevE.67.066203
[arXiv:cond-mat/0301273 [cond-mat]].
\bibitem{coscll}
G.~Miritello, A.~Pluchino and A.~Rapisarda,
``Phase Transitions and Chaos in Long-Range Models of Coupled Oscillators,''
EPL \textbf{85}, no.1, 10007 (2009)
doi:10.1209/0295-5075/85/10007
[arXiv:0807.1870 [cond-mat.stat-mech]].
\bibitem{finite}
W.~D.~Heiss and A.~L.~Sannino,
``Transitional regions of finite Fermi systems and quantum chaos,''
Phys. Rev. A \textbf{43}, 4159-4166 (1991)
doi:10.1103/PhysRevA.43.4159


\bibitem{static}
Y.~Sota, S.~Suzuki and K.~i.~Maeda,
``Chaos in static axisymmetric space-times. 1: Vacuum case,''
Class. Quant. Grav. \textbf{13}, 1241-1260 (1996)
doi:10.1088/0264-9381/13/5/034
[arXiv:gr-qc/9505036 [gr-qc]].
\bibitem{static2}
Y.~Sota, S.~Suzuki and K.~i.~Maeda,
``Chaos in static axisymmetric space-times. 2. Nonvacuum case,''
[arXiv:gr-qc/9610065 [gr-qc]].
\bibitem{kerr}
N.~Kan and B.~Gwak,
``Bound on the Lyapunov exponent in Kerr-Newman black holes via a charged particle,''
Phys. Rev. D \textbf{105}, no.2, 026006 (2022)
doi:10.1103/PhysRevD.105.026006
[arXiv:2109.07341 [gr-qc]].
\bibitem{kerr2}
B.~Gwak, N.~Kan, B.~H.~Lee and H.~Lee,
``Violation of bound on chaos for charged probe in Kerr-Newman-AdS black hole,''
JHEP \textbf{09}, 026 (2022)
doi:10.1007/JHEP09(2022)026
[arXiv:2203.07298 [gr-qc]].
\bibitem{multi}
W.~Hanan and E.~Radu,
``Chaotic motion in multi-black hole spacetimes and holographic screens,''
Mod. Phys. Lett. A \textbf{22}, 399-406 (2007)
doi:10.1142/S0217732307022815
[arXiv:gr-qc/0610119 [gr-qc]].
\bibitem{qg}
F.~Lu, J.~Tao and P.~Wang,
``Minimal Length Effects on Chaotic Motion of Particles around Black Hole Horizon,''
JCAP \textbf{12}, 036 (2018)
doi:10.1088/1475-7516/2018/12/036
[arXiv:1811.02140 [gr-qc]].
\bibitem{qg2}
X.~Guo, K.~Liang, B.~Mu, P.~Wang and M.~Yang,
``Chaotic Motion around a Black Hole under Minimal Length Effects,''
Eur. Phys. J. C \textbf{80}, no.8, 745 (2020)
doi:10.1140/epjc/s10052-020-8335-6
[arXiv:2002.05894 [gr-qc]].

\bibitem{mss}
J.~Maldacena, S.~H.~Shenker and D.~Stanford,
``A bound on chaos,''
JHEP \textbf{08}, 106 (2016)
doi:10.1007/JHEP08(2016)106
[arXiv:1503.01409 [hep-th]].
\bibitem{hori}
K.~Hashimoto and N.~Tanahashi,
``Universality in Chaos of Particle Motion near Black Hole Horizon,''
Phys. Rev. D \textbf{95}, no.2, 024007 (2017)
doi:10.1103/PhysRevD.95.024007
[arXiv:1610.06070 [hep-th]].
\bibitem{hori2}
S.~Dalui, B.~R.~Majhi and P.~Mishra,
``Presence of horizon makes particle motion chaotic,''
Phys. Lett. B \textbf{788}, 486-493 (2019)
doi:10.1016/j.physletb.2018.11.050
[arXiv:1803.06527 [gr-qc]].

\bibitem{vio}
Q.~Q.~Zhao, Y.~Z.~Li and H.~Lu,
``Static Equilibria of Charged Particles Around Charged Black Holes: Chaos Bound and Its Violations,''
Phys. Rev. D \textbf{98}, no.12, 124001 (2018)
doi:10.1103/PhysRevD.98.124001
[arXiv:1809.04616 [gr-qc]].
\bibitem{vio2}
X.~Guo, K.~Liang, B.~Mu, P.~Wang and M.~Yang,
``Minimal Length Effects on Motion of a Particle in Rindler Space,''
Chin. Phys. C \textbf{45}, no.2, 023115 (2021)
doi:10.1088/1674-1137/abcf20
[arXiv:2007.07744 [gr-qc]].
\bibitem{vio3}
B.~Gwak, N.~Kan, B.~H.~Lee and H.~Lee,
``Violation of bound on chaos for charged probe in Kerr-Newman-AdS black hole,''
JHEP \textbf{09}, 026 (2022)
doi:10.1007/JHEP09(2022)026
[arXiv:2203.07298 [gr-qc]].
\bibitem{vio4}
J.~Park and B.~Gwak,
``Bound on Lyapunov exponent in Kerr-Newman-de Sitter black holes by a charged particle,''
JHEP \textbf{04}, 023 (2024)
doi:10.1007/JHEP04(2024)023
[arXiv:2312.13075 [gr-qc]].
\bibitem{first}
X.~Guo, Y.~Lu, B.~Mu and P.~Wang,
``Probing phase structure of black holes with Lyapunov exponents,''
JHEP \textbf{08}, 153 (2022)
doi:10.1007/JHEP08(2022)153
[arXiv:2205.02122 [gr-qc]].
\bibitem{le}
S.~Yang, J.~Tao, B.~Mu and A.~He,
``Lyapunov exponents and phase transitions of Born-Infeld AdS black holes,''
JCAP \textbf{07}, 045 (2023)
doi:10.1088/1475-7516/2023/07/045
[arXiv:2304.01877 [gr-qc]].
\bibitem{le2}
X.~Lyu, J.~Tao and P.~Wang,
``Probing the thermodynamics of charged Gauss Bonnet AdS black holes with the Lyapunov exponent,''
Eur. Phys. J. C \textbf{84}, no.9, 974 (2024)
doi:10.1140/epjc/s10052-024-13354-9
[arXiv:2312.11912 [gr-qc]].
\bibitem{le3}
A.~N.~Kumara, S.~Punacha and M.~S.~Ali,
``Lyapunov exponents and phase structure of Lifshitz and hyperscaling violating black holes,''
JCAP \textbf{07}, 061 (2024)
doi:10.1088/1475-7516/2024/07/061
[arXiv:2401.05181 [gr-qc]].
\bibitem{le4}
Y.~Z.~Du, H.~F.~Li, Y.~B.~Ma and Q.~Gu,
``Phase structure and optical properties of the de Sitter Spacetime with KR field based on the Lyapunov exponent,''
Eur. Phys. J. C \textbf{85}, no.1, 78 (2025)
doi:10.1140/epjc/s10052-025-13809-7
[arXiv:2403.20083 [hep-th]].
\bibitem{le5}
N.~J.~Gogoi, S.~Acharjee and P.~Phukon,
``Lyapunov exponents and phase transition of Hayward AdS black hole,''
Eur. Phys. J. C \textbf{84}, no.11, 1144 (2024)
doi:10.1140/epjc/s10052-024-13520-z
[arXiv:2404.03947 [hep-th]].
\bibitem{le6}
B.~Shukla, P.~P.~Das, D.~Dudal and S.~Mahapatra,
``Interplay between the Lyapunov exponents and phase transitions of charged AdS black holes,''
Phys. Rev. D \textbf{110}, no.2, 024068 (2024)
doi:10.1103/PhysRevD.110.024068
[arXiv:2404.02095 [hep-th]].
\bibitem{le7}
D.~Chen, C.~Yang and Y.~Liu,
``Lyapunov exponents as probes for a phase transition of a Kerr-AdS black hole,''
Phys. Lett. B \textbf{865}, 139463 (2025)
doi:10.1016/j.physletb.2025.139463
[arXiv:2501.16999 [hep-th]].
\bibitem{le8}
K.~R., D.~D., K.~M.~Ajith, K.~Hegde, S.~Punacha and A.~N.~Kumara,
``Euclidean Thermodynamics and Lyapunov Exponents of Einstein-Power-Yang-Mills AdS Black Holes,''
[arXiv:2504.12890 [gr-qc]].
\bibitem{Awal}
M.~B.~Awal and P.~Phukon,
``Probing Thermodynamic Phase Transitions of 4D R-Charged Black Holes via Lyapunov Exponent,''
[arXiv:2505.20800 [gr-qc]].
\bibitem{le9}
C.~Yang, C.~Gao, D.~Chen and X.~Zeng,
``Lyapunov exponents, phase transition and chaos bound in Kerr-Newman AdS spacetime,''
[arXiv:2506.21882 [hep-th]].
\bibitem{le10}
X.~Guo, R.~Yang, Y.~Liang and J.~Tao,
``Lyapunov exponents and phase transition of charged Ads black hole in quintessence,''
[arXiv:2508.03519 [gr-qc]].


\bibitem{mm}
I.~Bandos, K.~Lechner, D.~Sorokin and P.~K.~Townsend,
``A non-linear duality-invariant conformal extension of Maxwell's equations,''
Phys. Rev. D \textbf{102} (2020), 121703
doi:10.1103/PhysRevD.102.121703
[arXiv:2007.09092 [hep-th]].

\bibitem{Sekhmani}
Y.~Sekhmani, S.~K.~Maurya, M.~K.~Jasim, {\.I}.~Sakall{\i}, J.~Rayimbaev and I.~Ibragimov,
``Thermodynamics and phase transition of anti de Sitter black holes with ModMax nonlinear electrodynamics and perfect fluid dark matter,''
Eur. Phys. J. C \textbf{85} (2025) no.3, 229
doi:10.1140/epjc/s10052-025-13932-5
\bibitem{Kosyakov}
B.~P.~Kosyakov,
``Nonlinear electrodynamics with the maximum allowable symmetries,''
Phys. Lett. B \textbf{810} (2020), 135840
doi:10.1016/j.physletb.2020.135840
[arXiv:2007.13878 [hep-th]].

\bibitem{ce}
R.~Banerjee and D.~Roychowdhury,
``Critical behavior of Born Infeld AdS black holes in higher dimensions,''
Phys. Rev. D \textbf{85} (2012), 104043
doi:10.1103/PhysRevD.85.104043
[arXiv:1203.0118 [gr-qc]].

\bibitem{Hashimoto}
K.~Hashimoto and N.~Tanahashi,
``Universality in Chaos of Particle Motion near Black Hole Horizon,''
Phys. Rev. D \textbf{95} (2017) no.2, 024007
doi:10.1103/PhysRevD.95.024007
[arXiv:1610.06070 [hep-th]].

\bibitem{Lei}
Y.~Q.~Lei, X.~H.~Ge and S.~Dalui,
``Thermodynamic stability versus chaos bound violation in D-dimensional RN black holes: Angular momentum effects and phase transitions,''
Phys. Lett. B \textbf{856}, 138929 (2024)
doi:10.1016/j.physletb.2024.138929
[arXiv:2404.18193 [hep-th]].








\end{thebibliography}
\end{document}